\documentclass{aa}
\usepackage{graphicx}
\begin{document}
   \title{A molecular line survey of the candidate massive Class 0 protostar IRAS
   23385+6053}

   \titlerunning{A line survey of IRAS 23385+6053}
   \author{M.A.~Thompson
          \and
          G.H.~Macdonald}

   \offprints{M.A.~Thompson}

   \institute{Centre for Astrophysics \& Planetary Science,
              School of Physical Sciences,
              University of Kent,
              Canterbury,
              Kent CT2 7NR,
              UK\\
              \email{m.a.thompson@kent.ac.uk, g.h.macdonald@kent.ac.uk}
	     }

   \date{Received 8 April 2003 / Accepted 21 May 2003}

   \abstract{We have carried out a molecular line survey of the candidate massive
   protostar \object{IRAS 23385+6053}, covering a 27.2 GHz frequency range in the 330--360
   GHz atmospheric window. We detected 27  lines originating from a total of 11 species.
   Over a third of the identified molecular lines are from the asymmetric top molecule
   methanol (CH$_{3}$OH). We did not detect any emission from high-excitation lines or
   typical hot core tracers (e.g.~CH$_{3}$CN, HCOOCH$_{3}$). We derive a rotation
   temperature and column density from the methanol emission and estimate lower limits to
   the beam-averaged  column density of the remaining lines. Upper limits to the
   beam-averaged column density of selected species were determined from the non-detection
   of their rotation lines. We rule out the presence of a hot molecular core associated
   with IRAS 23385+6053 by a combination of the non-detection of CH$_{3}$CN emission and a
   simple bolometric luminosity approach. The molecular inventory and chemistry of IRAS
   23385+6053 are contrasted to that of more evolved massive star-forming regions and the
   abundances predicted by recent time-dependent chemical models. The physical and
   chemical nature of \object{IRAS 23385+6053} is shown to be consistent with that
   immediately prior to the hot molecular core stage.   }

   \maketitle
%


\newcommand{\aap}{A\&A}
\newcommand{\aaps}{A\&AS}
\newcommand{\apj}{ApJ}
\newcommand{\apjs}{ApJS}
\newcommand{\mnras}{MNRAS}
\newcommand{\araa}{ARA\&A}
\newcommand{\msol}{$M_{\odot}$}
\newcommand{\lsol}{$L_{\odot}$}
\newcommand{\tastar}{$T_{A}^{*}$}
\newcommand{\trstar}{$T_{R}^{*}$}
\newcommand{\etafss}{$\eta_{\rm fss}$}


\section{Introduction}

The evolutionary process in which a high-mass star develops within a molecular cloud
core is still subject to some uncertainty. Low-mass star formation, on the other hand, 
is underpinned by an extremely successful classification system (e.g.~Andr\'e et
al.~\cite{awb00}), stretching from the earliest protostellar phase (Class 0) to 
obscured young stellar objects and T-Tauri stars (Class II/III). No analogous scheme 
exists for high-mass protostars and Young Stellar Objects (YSOs), although a tentative
scheme linking the development of hot molecular cores to that of ultracompact HII
regions has recently been proposed (Kurtz et al.~\cite{kcchw00}).  In part this is due
to the observational challenges posed by objects that are less numerous, are at much
greater distances, preferentially form in compact clusters and evolve more rapidly than
low-mass star-forming regions. 

Until recently, the earliest known phase in the development of a massive YSO was the
ultracompact HII region (Churchwell \cite{churchwell02}), which traces the development of a
high-mass star some 10$^{5}$ years after its formation (De Pree et al.~\cite{dprg95}). Much
effort in recent years has been placed into identifying the earlier phases of high mass
star formation, i.e.~the massive protostellar precursors of ultracompact HII regions (e.g.
Molinari et al.~\cite{mol96}, \cite{mol98a}, \cite{mol00}; Sridharan et al.~\cite{sbsmw02},
Beuther et al.~\cite{bsmmsw02}, Lumsden et al.~\cite{lhor02}). A large number of candidate
massive protostars and YSOs have been identified in these studies  and the physical
conditions of the pre-ultracompact HII region phase are beginning to emerge.

One of the prototypes of these candidate massive protostars is \object{IRAS 23385+6053},
identified in the search of Molinari et al.~(\cite{mol96}, \cite{mol98a}). It is comprised
of a compact dense molecular core,  massing some 370 \msol, associated with a powerful
molecular outflow seen in HCO$^{+}$ and SiO emission (Molinari et al.~\cite{mol98b}).
Although there is nearby extended radio and mid-infrared emission (Molinari et
al.~\cite{mol98a}, \cite{mol98b}) associated with two YSO clusters, the dense molecular
core of IRAS 23385+6053 is coincident with neither (see Fig.~\ref{fig:msxfig}) and is 
detected only by its far-infrared to sub-millimetre continuum emission. Molinari et
al.~(\cite{mol98b}) determined the sub-millimetre to bolometric luminosity ($L_{\rm
submm}/L_{\rm bol}$) and envelope mass to core mass ($M_{\rm env}/M_{*}$) ratios for
\object{IRAS 23385+6053}, which are standard diagnostic indicators for the evolutionary
class of low-mass protostars (Andr\'e et al.~\cite{awb93}). Based on these indicators and
the large bolometric luminosity  ($\sim1.6\times10^{4}$ \lsol\ for an assumed kinematic
distance of 4.9 kpc) Molinari et al.~(\cite{mol98b}) concluded that \object{IRAS
23385+6053} was the first bona fide example of a massive class 0 protostar.

Although a wealth of information on the physical properties of massive protostars and YSOs
is becoming available there is as yet little information on their chemistry or molecular
inventory. The gas-phase molecular chemistry of star-forming regions has long been regarded
as a potential diagnostic of their age (e.g.~Brown et al.~\cite{bcm88}). Many chemical
studies have been made of hot molecular cores and ultracompact HII region (Gibb et
al.~\cite{gniwb00}; Thompson et al.~\cite{tm99}; Hatchell et al.~\cite{htmm98a},
\cite{htmm98b}, Schilke et al.~\cite{sgbp97}), but none to date of candidate massive
protostellar objects. Molecular line surveys allow the abundances of a large number of
molecular species to be determined, characterising the chemistry of the molecular gas and
providing a list of molecular tracers that may be used to probe the physical conditions of
massive protostars (e.g.~to determine the gas kinetic temperature or density, or for use in
infall searches). For these reasons we have carried out a molecular line survey of the
candidate massive protostar IRAS 23385+6053 with the James Clerk Maxwell Telescope.

The observations and data reduction procedure are detailed in the next section. The
identified molecular lines detected in the survey and their derived column densities and
abundances are presented in Sect.~\ref{sect:analysis}. The results of the survey and their
implications for the protostellar nature of \object{IRAS 23385+6053} and its relation to the more
evolved hot molecular cores and ultracompact HII regions are discussed in
Sect.~\ref{sect:discuss}.


\section{Observations and data reduction}

\subsection{Observations}

The observations were carried out at the James Clerk Maxwell Telescope (JCMT\footnote{
The JCMT is operated by the Joint Astronomy Centre on behalf of PPARC for the United
Kingdom, the Netherlands Organisation of Scientific Research, and the National Research
Council of Canada.}) during the August 1999 to January 2000 semester. The observations
were scheduled according to the flexible-scheduling backup mode in operation at the
JCMT, whereby scientific programmes are accorded a priority within a specified weather
band defined by the atmospheric opacity at 225 GHz and then carried out by visiting
observers according to suitable weather conditions. The data were thus obtained
piecemeal over a period of several months, depending upon the local weather conditions
and the visibility of the target source \object{IRAS 23385+6053}.

\begin{figure}
\centering
\includegraphics*[scale=0.39,trim=130 20 150 70]{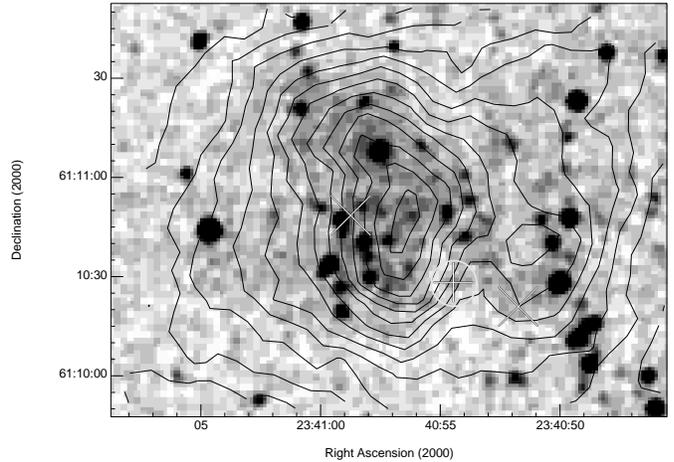}
\caption{2MASS $K_{s}$ band image of the region surrounding \object{IRAS 23385+6053}. The contours
represent the 8 $\mu$m emission measured by MSX, with a starting contour level of $10^{-6}$
W\,m$^{-2}$\,sr$^{-1}$ and contour spacing of $10^{-6}$
W\,m$^{-2}$\,sr$^{-1}$. The position of
the millimetre peak identified by Molinari et al.(\cite{mol98b}) is marked with a
vertical cross (+) and the two adjacent radio sources \object{Mol 160 VLA 1} and \object{Mol
160 VLA 2} are shown by
diagonal crosses ($\times$). The molecular line survey position and beam FWHM is
indicated by a white circle.}
\label{fig:msxfig}
\end{figure}

The 330--360 GHz region was selected for the molecular line survey to match as closely
as possible the FWHM beam-width of the JCMT at this frequency (14\arcsec\ at 345 GHz) to
the observed angular diameter of the \object{IRAS 23385+6053} molecular core (FWHM $\sim
5$\arcsec) as measured in the HCO$^{+}$ J=1--0 transition, Molinari et
al.~\cite{mol98b}). This approach was crucial to minimise the beam dilution caused by
observing such a compact source and also to avoid possible contamination from gas associated
with the nearby mid-infrared sources (see Fig.~\ref{fig:msxfig}). A secondary concern
for choosing this frequency range was to facilitate the comparison of the survey results
with the many surveys of hot cores and ultracompact HII regions carried out in this
range (Thompson et al.~\cite{tm99}; Schilke et al.~\cite{sgbp97};  Macdonald et
al.~\cite{mghm96}, Jewell et al.~\cite{jhls89}).

All observations were made at the coordinates of the millimetre peak of \object{IRAS 23385+6053}
given by Molinari et al.~(\cite{mol98b}), i.e. $\alpha$(2000) = 23$^{\rm h}$40$^{\rm
m}$54\fs 5, $\delta$(2000) = +61\degr 10\arcmin 28\arcsec . The pointing accuracy of the
telescope was checked hourly against standard pointing calibrators and was found to be
good to within 3--4\arcsec. The spectra were taken in beam-switching mode, in which the
secondary mirror is chopped from on-source to off-source at a frequency of 1 Hz.
Beam-switching is much superior to position-switching (i.e.~moving the primary mirror to
the off-source position) for obtaining the extremely flat baselines required to search
for faint line emission. A chop throw of 3\arcmin\ in RA was used to keep a constant
reference position from spectrum to spectrum. This throw was more than sufficient to
reach a clean reference position for all species except CO, which  in massive
star-forming regions is typically widespread over a much larger area. 

The spectra were observed using the facility heterodyne receiver B3 (RxB3), which is a dual-SIS
junction receiver operating in the 345 GHz band, coupled to the facility Digital
Autocorrelation Spectrometer (DAS). The receiver was used in dual sideband
mode in which the resulting spectra are comprised of two frequency bands (the upper and
lower sidebands)  folded over one another. The ``main band'' of the spectrum may be set
to either the upper or lower sideband and the remaining sideband is often referred to as
the ``image band''. The main and image bands are separated from one another by roughly
twice the intermediate frequency or IF of the receiver, which for RxB3 is 4 GHz,
leading to a band separation of $\sim$ 8 GHz. 

The receiver may also be used in single-sideband mode where the image band of the
receiver is attenuated by roughly  a factor of $\sim$ 20 using a dual-beam Mach-Zender
interferometer. Strong lines in the image band may thus leak through to the main band, 
mimicking faint main band lines and the sideband rejection factor is not accurately
calibrated over the entire passband of the receiver. We used the dual-sideband
mode of receiver B3  to avoid problems in relative image band rejection across the
passband and to instantaneously obtain twice the frequency coverage of single-sideband
mode. The sideband of each detected line was determined by taking an additional spectrum
with the local oscillator frequency shifted by +10 MHz. In the shifted spectrum lines in
the image band appear to shift frequency by $\pm20$ MHz relative to the lines in the main
band.

The DAS was set to a bandwidth of 920 MHz to exploit the maximum dual-sideband bandwidth
of RxB3. Our observing strategy was to take spectra with the main band set to the
lower sideband, increment the central frequency of the main band by 800 MHz and repeat
the spectra until the frequency range covered in the lower sideband reached that of the
upper sideband from the first spectrum. A  block of 10 spectra observed in this manner
sample contiguous 8 GHz frequency ranges in both the lower and upper sidebands. Two of
these blocks of dual-sideband spectra are sufficient to cover the entire 330--360 GHz
frequency range, with a 1 GHz extension to either side. The 800 MHz frequency increment allows for a 120 MHz overlap between
spectra, allowing the  the raised noise at the edge of the passband to be avoided.
Unfortunately, during the observations the second block of spectra was mistakenly
observed in the the upper sideband rather than the lower sideband, leading to a
duplication of the 338--342 GHz frequency range and a gap in the coverage between
352--359 GHz.

\begin{figure}[h]
\includegraphics[angle=-90,scale=0.6]{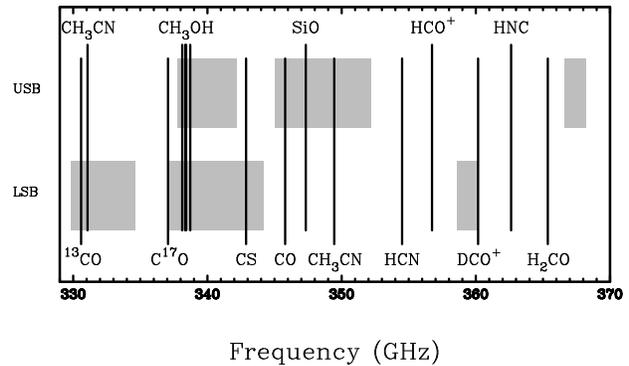}
\caption{Spectral coverage of the survey. Shaded boxes indicate
 the frequency regions covered in
the Upper (USB) and Lower (LSB) sidebands of the receiver. The frequencies of various
important astrophysical transitions are marked.}
\label{fig:coverage}
\end{figure}

An attempt was made to take extra spectra to cover the gaps in frequency coverage,
although the lack of suitable weather over the semester meant that the gaps could not be
completely filled. The resulting frequency coverage of the survey is 27.2 GHz with gaps
at 334.6--337, 344.2--345 and 352-358.6 GHz. Fig.~\ref{fig:coverage} shows the total
frequency coverage of the survey. Despite the gaps in frequency coverage the molecular
line survey is 85\% complete over the 329--361 GHz range originally intended and covers
the frequencies of many important astrophysical transitions. The implications of the
gaps in frequency coverage and their impact on the survey are discussed further in
Sect.~\ref{sect:discuss}.

With the DAS set to a 920 MHz bandwidth the width of each channel was 625 kHz, with a
spectral resolution of 756 kHz. Later in the data reduction procedure all spectra were
binned to a channel width of 1.25 MHz to improve the signal to noise ratio. The standard
three-load chopper-wheel calibration method of Kutner \& Ulich (\cite{ku81}) was used to
obtain line temperatures on the \tastar\ scale, i.e.~corrected for the atmosphere,
resistive telecope losses and rearward spillover and scattering. All line temperatures
quoted in this paper are on the \tastar\ scale unless explicitly stated otherwise. Values
of \tastar\ may be converted to the corrected receiver temperature \trstar\ by dividing
by the forward spillover and scattering coefficient \etafss, which for the JCMT and RxB3
is 0.7 at 345 GHz. Absolute calibration of the \tastar\ scale was determined by regular
observations of the standard spectral line calibrator NGC 7538 IRS1 and was found to be
accurate to within 10\%.

\subsection{Data reduction and line identification}

\begin{figure}
\centering
\includegraphics*[angle=-90,scale=0.3]{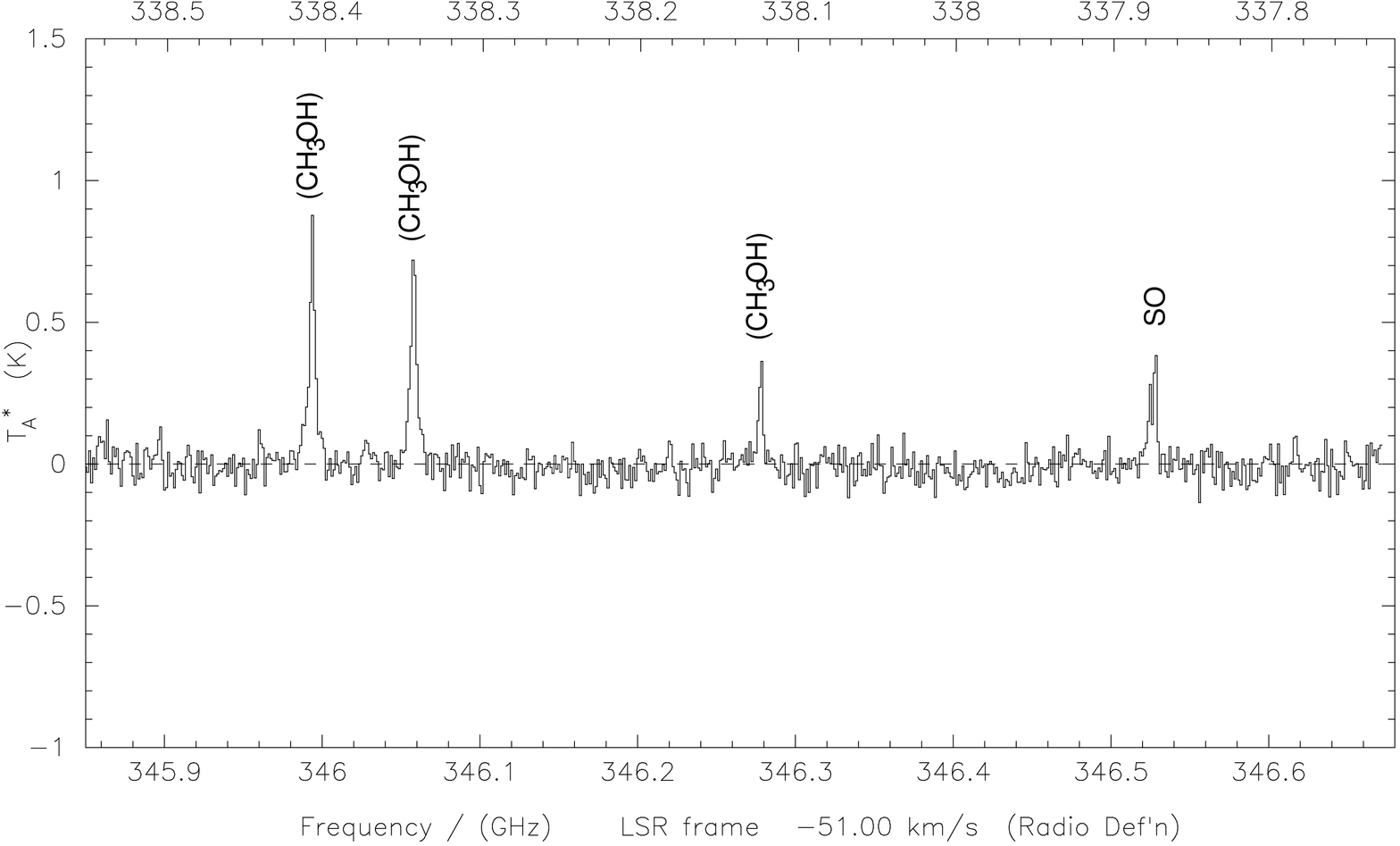}
\includegraphics*[angle=-90,scale=0.3,trim=-30 0 0 0]{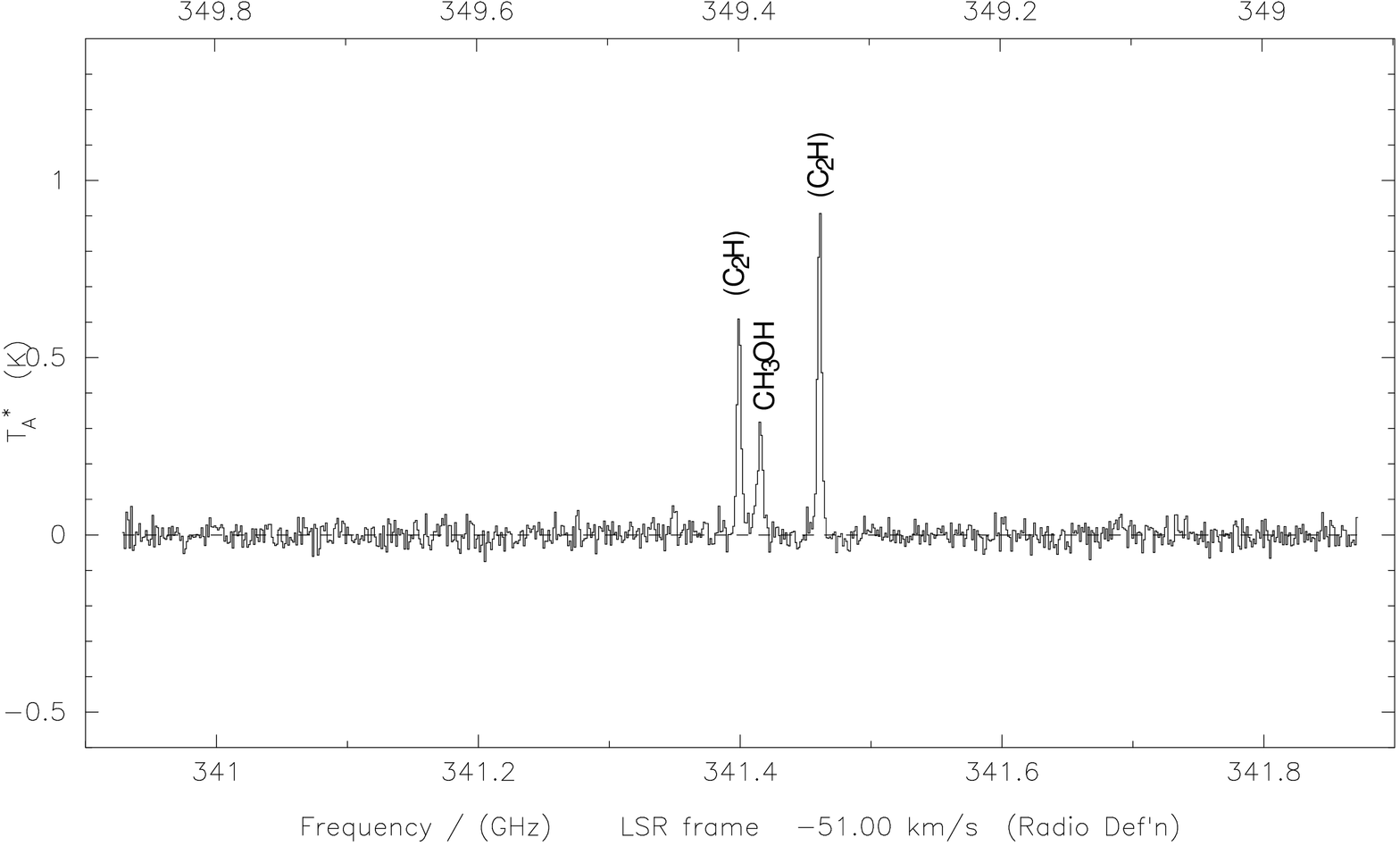}
\caption{Sample dual sideband spectra from the survey. The main band frequency scale is
shown on the bottom axis and the image band scale on the top axis of each spectrum. 
Identified lines are marked by
their molecule name and lines in the image band are indicated by brackets.}
\label{fig:spectra}
\end{figure}

The data were reduced using the Starlink millimetre-wave spectroscopy package SPECX
(Prestage et al.~\cite{specx}). Linear baselines were subtracted from each
spectrum and the line parameters of peak temperature (\tastar), central observed
frequency ($\nu$(obs)) and line width at half maximum ($\Delta\nu_{1/2}$) were measured.
A value of $-$51.0 km\,s$^{-1}$ for the LSR velocity of \object{IRAS 23385+6053} was assumed,
following Molinari et al.~(\cite{mol98b}). Sample spectra are shown in Figure
\ref{fig:spectra}.

The noise in each spectrum
was estimated using the line-free channels and was found to be typically 50 mK per 1.25
MHz wide channel. Line features below a minimum detection limit of 3$\sigma$ (0.15 K)
were ignored to avoid inaccurate line identifications. Each line was identified at least
twice in separate spectra, including the 10MHz shifted spectra and the 120 MHz overlaps
between adjacent spectra. The values of \tastar, $\nu$(obs) and $\Delta\nu_{1/2}$ were
averaged together for multiple detections of the same line and these average quantities
are listed in Table \ref{tbl:linetable}.

\begin{table*}[ht]
\caption{The measured line parameters of observed frequency ($\nu$(obs)), peak
antenna temperature ($T_{A}^{*}$) and line width ($\Delta\nu_{1/2}$) for each line
detected in the survey are listed here. Note that at 345 GHz a line width of 1 MHz is
equivalent to 0.87 km\,s$^{-1}$. The line parameters of 
multiply detected lines have been averaged together. Possible other species or
hyperfine blends with the detected lines are indicated in the Notes column. Where a
single line may be identified with one or more species both possibilities are listed
with the most likely indicated by an asterisk (*). The notation \emph{sl-blend}
refers to blended lines that may have one or both components extracted.}

\label{tbl:linetable}
\begin{tabular}{lccccccl}\hline
$\nu$(obs) & $T_{A}^{*}$ & $\Delta\nu_{1/2}$ & Species & Transition & $\nu$(rest) & $E_{\rm u}/k$ & Notes \\
 (GHz) & (K) & (MHz) & & & (GHz) & (K) & \\ \hline
330.587 & 6.89 & 4.4 & $^{13}$CO & 3--2 & 330.588 & 31.7 & \\
337.060 & 0.90 & 2.7 & C$^{17}$O & 3--2 & 337.061 & 32.4 & \\
338.124 & 0.33 & 3.3 & CH$_{3}$OH & 7(0)--6(0) E & 338.125 & 76.9 & \\
338.344 & 0.73 & 4.6 & CH$_{3}$OH & 7($-$1)--6($-$1) E & 338.345 & 69.4 & \\
338.408 & 0.89 & 4.6 & CH$_{3}$OH & 7(0)--6(0) A+ & 338.409 & 65.0 & \\
338.515 & 0.21 & 3.9 & CH$_{3}$OH & 7(2)--6(2) A$-$ & 338.513 & 103 & \\
338.615 & 0.22 & 3.9 & CH$_{3}$OH & 7(1)--6(1) E & 338.615 & 84.9 & \\
338.722 & 0.27 & 3.6 & CH$_{3}$OH & 7(2)--6(2) E & 338.722 & 86.1 & \\
340.032 & 0.25 & 2.8 & CN & 3--2 2.5 3.5--1.5 2.5 & 340.032 & 16.3 & sl-blend\\ 
340.035 & 0.24 & 2.8 & CN & 3--2 2.5 1.5--1.5 0.5 & 340.035 & 16.3 & h/fine blend with 2.5
2.5-1.5 1.5\\
340.247 & 0.59 & 3.5 & CN & 3--2 3.5 4.5--2.5 3.5 & 340.248 & 16.3 & \\
340.714 & 0.26 & 3.8 & SO & 7(8)--6(7) & 340.714 & 81.2 & \\
341.350 & 0.15 & 2.4 & HCS$^{+}$ & 8--7 & 341.350 & 73.7 & \\
341.415 & 0.38 & 4.2 & CH$_{3}$OH & 7(1)--6(1) A$-$ & 341.416 & 80.1 & \\
342.883 & 1.25 & 3.8 & CS & 7--6 & 342.883 & 65.8 & \\
345.339 & 0.15 & 3.7 & H$^{13}$CN & 4--3 & 345.340 & 41.3 & * \\
& & & SO$_{2}$ & 13(2,12)--12(2,11) & 345.339 & 93.0 & \\
345.795 & 12.19 & -- & CO & 3--2 & 345.796 & 33.2 & strongly self-absorbed \\
346.527 & 0.35 & 4.7 & SO & 9(8)--8(7) & 346.529 & 78.8 & \\
346.998 & 0.19 & 4.4 & H$^{13}$CO$^{+}$ & 4--3 & 346.999 & 41.6 & \\
347.387 & 0.17 & 5.0 & U & & & & \\
348.534 & 0.15 & 5.0 & H$_{2}$CS & 10(1,9)--9(1,8) & 348.532 & 105 & \\
349.338 & 0.92 & 4.9 & C$_{2}$H & 4.5--3.5 & 349.338 & 41.9 & \\
349.400 & 0.61 & 3.7 & C$_{2}$H & 3.5--2.5 & 349.401 & 41.9 & \\
350.688 & 0.47 & 4.4 & CH$_{3}$OH & 4(0)--3($-$1) E & 350.688 & 35.1  & poss.~blend with
NO 3.5 0.5--2.5 0.5\\
350.905 & 0.38 & 4.9 & CH$_{3}$OH & 1(1)--0(0) A+ & 350.905 & 16.8 & \\
351.768 & 1.60 & 4.9 & H$_{2}$CO & 5(1,5)--4(1,4) & 351.769 & 62.5 & \\
358.606 & 0.23 & 3.9 & CH$_{3}$OH & 4(1)--3(0) E & 358.606 & 43.1 & \\
\hline\end{tabular}
\end{table*} 

The lines were identified with molecular transitions by comparing their central observed
frequencies to those listed in the JPL Molecular Spectroscopy Database (available from
\texttt{http://spec.jpl.nasa.gov}). Other line lists used include  
the methanol lists of Anderson et al.~(\cite{anderson}) and the observational lists of
lines detected by Jewell et al.~(\cite{jhls89}) toward Orion-A, Macdonald et
al.~(\cite{mghm96}) toward G34.3+0.15, Schilke et al.~(\cite{sgbp97}) toward Orion-KL and
Thompson \& Macdonald (\cite{tm99}) toward G5.89$-$0.39.  A list of species
identified in the molecular gas of \object{IRAS 23385+6053} may be found in Table
\ref{tbl:speciestable}. The identification of asymmetric rotors (such as H$_{2}$CS) from
single line detections must be viewed with caution as these species possess many other
transitions within the observed frequency range.

\begin{table}[h]
\caption{Molecular species (including isotopomers) identified in the survey. We note that the identification
of certain species (e.g.~H$_{2}$CS) from single line detections must be regarded with
caution.}
\label{tbl:speciestable}
\begin{tabular}{lc}\hline
Species & Number of \\
 & detected lines \\ \hline\
CO & 1 \\
$^{13}$CO & 1 \\
C$^{17}$O & 1 \\
CH$_{3}$OH & 10 \\
CN & 3 \\
SO & 2 \\
HCS$^{+}$ & 1 \\
CS & 1 \\
H$^{13}$CN & 1 \\
H$^{13}$CO$^{+}$ & 1 \\
H$_{2}$CS & 1 \\
C$_{2}$H & 2 \\
H$_{2}$CO & 1 \\ \hline
\end{tabular}
\end{table}
 

\section{Results and analysis}
\label{sect:analysis}

\subsection{Identified molecular lines}

A total of 27 lines were identified in the survey, originating from 11 species, and are
shown in Table \ref{tbl:linetable}. One further line at 347.387 GHz  could not be
identified with any  transition listed in the JPL Molecular Spectroscopy Database and is
also not listed in any of the other observational lists that were checked. A list of the
identified species (including isotopomers) and the number of detected lines from each is 
contained in Table \ref{tbl:speciestable}. Most of the lines identified in the survey
originate from the J=7--6 transitions of methanol (CH$_{3}$OH). With the exception of the
$^{12}$CO J=3--2 line the lines are narrow, with typical FWHM linewidths of $\sim$ 4 MHz,
which corresponds to a width of 3.5 km\,s$^{-1}$ at 345 GHz.

Emission from  the commonly used tracers of temperature and column density
CH$_{3}$CN (methyl cyanide) and CH$_{3}$CCH (propyne or methyl acetylene) was not detected 
down to a level of $\sim 0.15$ K. If the most
likely identification of the line at 345.399 GHz is the J=4--3 transition of H$^{13}$CN
then the SO$_{2}$ molecule was also not detected toward \object{IRAS 23385+6053}. 
No transitions from the organic asymmetric rotors
(e.g.~CH$_{3}$OCH$_{3}$, HCOOCH$_{3}$, NH$_{2}$CHO, H$_{2}$CCHCN or C$_{2}$H$_{5}$OH)
observed to be widespread in hot molecular cores (e.g.~Hatchell et al.\cite{htmm98a}) 
were detected.

The non-detections of these various transitions most likely arise from  a mix of
excitation and chemical effects. The narrow linewidths and faint molecular lines
observed in the survey indicate that the molecular gas of \object{IRAS 23385+6053} is cold,
meaning that there is an insufficient kinetic temperature to adequately  excite the
higher energy transitions. High excitation lines require hot gas in order to be excited.
The highest excitation transition observed in the survey was the 7(2)--6(2) A$-$
CH$_{3}$OH line at 338.515 GHz, which possesses an upper energy level of $E_{\rm u}/k =
103$ K. It is likely that the non-detection of certain species whose lines in this
frequency range are of high excitation (e.g.~HC$_{3}$N with an $E_{\rm u}$/k $\sim$ 300
K) is reflected by the inability of the cold gas to excite their transitions adequately
rather than a genuinely low gas-phase abundance. However, asymmetric rotors such as
SO$_{2}$, HCOOCH$_{3}$ and C$_{2}$H$_{5}$OH possess many lines across the frequency
range of the survey with a wide range of excitation. The non-detection of the
low-excitation lines of these species must arise from their low gas-phase abundance
rather than excitation effects.  We will further explore these issues and their
implications for the chemistry of \object{IRAS 23385+6053}  in Sect.~\ref{sect:discuss}.

\subsection{Rotation diagram analysis}

The temperature and column density of the gas may be determined via rotation diagrams
(Turner \cite{turner91}),
also known as Boltzmann plots (e.g. Brand et al.~\cite{bcpm01}). The main
assumptions in the rotation diagram approach are that the gas is optically thin, 
in Local Thermodynamic
Equilibrium (LTE) and can be described by a single rotational temperature $T_{\rm
rot}$. For these assumptions the column density equation may be written as
(e.g.~Thompson, Macdonald \& Millar \cite{tmm99})

\begin{equation}
\label{eqn:coldensity}
N = \frac{3k}{8\pi^{3}}\, \frac{\int T_{R} \,\,{\rm d}v}{\nu S \mu^{2} g_{I}g_{K}} \,Q(T_{\rm
rot})\,{\rm exp}\left(\frac{E_{\rm u}}{k T_{\rm rot}}\right), 
\end{equation}

\noindent where $\int T_{R} \,\,{\rm d}v$ is the integrated intensity of the line, $\nu$
is the line frequency, $S$ is the line strength, $\mu$ is the permanent dipole moment,
$g_{I}$ and $g_{K}$ are, respectively, the reduced nuclear spin and K-level degeneracy
of the molecule. The energy of the upper level of the line is represented by $E_{\rm
u}$. The partition function as a function of $T_{\rm rot}$ is written as $Q(T_{\rm
rot})$ and represents the partitioning of the total level energies into each rotational
level. 

The column density $N$ derived from Eq.~\ref{eqn:coldensity} 
is beam-averaged rather than source-averaged, as the beam-averaged receiver temperature
\trstar\ is used instead of the source brightness temperature $T_{\rm b}$. To calculate a
source-averaged column density the beam-filling factor of the emission (and hence the
emitting area of the line) must be known. Each line will trace slightly different
environments depending upon their critical densities and excitation temperature, thus
there is no guarantee that the beam filling factors are constant for all species and
lines. As the beam filling factors are unknown for the transitions in question we derive
beam-averaged column densities throughout this paper. 

Equation
\ref{eqn:coldensity} is often known as the column density equation and, by taking
logarithms of each side, it is possible to rearrange this equation to that of a straight
line, i.e.

\begin{equation}
\label{eqn:rotdiag}
\log \left(\frac{3k}{8\pi^{3}} \frac{\int T_{R} \,\,{\rm d}v}{\nu S \mu^{2} g_{I}g_{K}} \right)
 = \log\left(\frac{N}{Q(T_{\rm rot})}\right) - \frac{E_{\rm u}}{k}\,\frac{\log e}{T_{\rm
 rot}}
\end{equation}

\noindent This equation is more commonly known as the rotation diagram equation
(e.g.~Macdonald et al.~\cite{mghm96}). The left hand quantity in Equation
\ref{eqn:rotdiag} is more commonly known as $L$ and a plot of $L$ against
$E_{\rm u}/k$ results in a straight line with a gradient of $-log\,\,e/T_{\rm rot}$ and
an intercept of $\log\left(N/Q(T_{\rm rot})\right)$. Methanol is the only species in our
survey with a sufficient number of identified lines to plot rotation diagrams. Methanol
is an asymmetric top molecule with its rotation-torsion levels split into two
types of sublevel: A (symmetric) and E (degenerate). Both types must be analysed
separately to take into account differences in sublevel populations.

\begin{figure}
\centering
\includegraphics*[trim=60 0 60 0,scale=0.4]{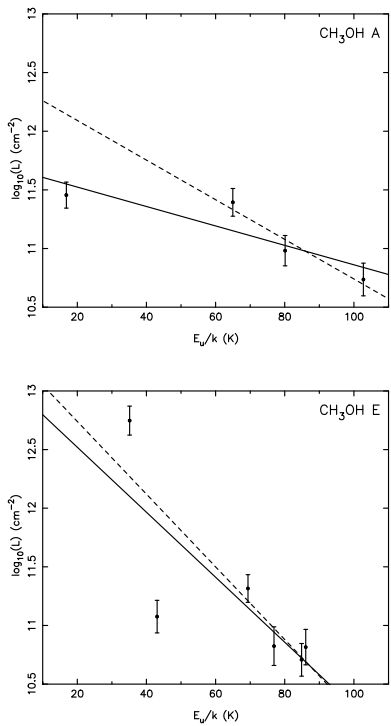}
\caption{Rotation diagrams of the A and E forms of methanol. Solid lines represent linear
least-squares fits to all the data points, whereas dashed lines are least-squares fits to
lines with $E_{\rm u}/k < 50$ K (as mentioned in  the text).}
\label{fig:meth_rotdiag}
\end{figure}

We plotted separate rotation diagrams for the A and E-type methanol transitions identified
in the survey, which may be seen in Fig.~\ref{fig:meth_rotdiag}. Straight lines were
fitted to the data using a least-squares fitting routine. The results of the straight line
fits for the A-type transitions are $T_{\rm rot} = 52 \pm 20$ K and $N = 2.4 \pm
1.5\,\,10^{14}$ cm$^{-2}$. The fits to the E-type transitions yield a rotation temperature
and column density of $T_{\rm rot} = 16 \pm 6$ K and $N = 9.4 \pm 8.9\,\,10^{14}$
cm$^{-2}$. The partition function $Q(T_{\rm rot})$ was evaluated  via interpolation
from the values quoted in the  JPL Molecular Spectroscopy Database to the rotation
temperature $T_{\rm rot}$ derived from the rotation diagram. The column densities were
then  determined using Eq.~\ref{eqn:rotdiag} and the y-intercept value from the
least-squares fit.

What can be immediately seen from the two rotation diagrams in Figure
\ref{fig:meth_rotdiag} is that the lower excitation transitions are inconsistent with the
straight line fits. When transitions with values of $E_{\rm u}/k$ less than 50 K are
excluded from the straight line fits then the straight line correlation coefficient
improves markedly for both rotation diagrams. We carefully inspected the data for any
irregularities (e.g.~pointing offsets, line misidentifications or blending, calibration
errors)  which may have caused the integrated intensity or upper level energy of any of
the methanol lines to be over- or under-estimated. No such cause was found and we conclude
that  it is likely to be a genuine physical effect rather than an artefact of the
observations. Discounting the low excitation lines with $E_{\rm u}/k \leq 50$ K from the
straight line fit brings the rotation temperatures and column densities derived from each
subtype into closer agreement, with $T_{\rm rot} \simeq 20$ K and $N \simeq 10^{15}$
cm$^{-2}$. These straight line fits are indicated in Fig.~\ref{fig:meth_rotdiag} by
dashed lines.

The most likely cause of the difference in the low-excitation transitions is a breakdown
in the assumptions behind the rotation diagram approach, i.e.~the gas may not be
described by a single temperature or it is optically thick. If the former were true then
the low excitation lines would predominantly originate from colder gas than the high
excitation lines and one would expect the low-excitation lines to follow a steeper
(i.e.~colder) gradient. If the latter were true then the integrated intensity of the 
more optically thick lower excitation lines would be underestimated compared with the less
optically thick higher excitation lines. 

It is unlikely that the methanol lines are
appreciably optically thick, as they are relatively faint and 
show no evidence of the line core saturation or asymmetric line profiles expected for
optically thick lines. We do not detect the low excitation lines of the 
$^{13}$CH$_{3}$OH isotopomer, although the upper limit on their detection (\tastar $\leq
0.15$ K) does not allow stringent limits to be placed on the optical depth of the low
excitation lines ($\tau_{\rm max} \leq 20$). There is some evidence for a two-temperature
distribution in the E-type rotation diagram, as the two lowest excitation lines appear
to follow a steeper gradient. Further observations of lower excitation lines, for
example in the 210--280 GHz window, are required to confirm this hypothesis.

\subsection{Lower limits to column density}

Eq.~\ref{eqn:coldensity} may be used to yield lower limits to the beam-averaged
column density for species with only one or two detected lines. To derive a lower limit
to the beam-averaged column density we set the derivative of the temperature-dependent part of
Eq.~\ref{eqn:coldensity}  to zero, i.e.

\begin{equation}
\frac{\rm d}{{\rm d}T_{\rm ex}}\left[Q(T_{\rm ex})\,\exp(E_{\rm u}/kT_{\rm ex})\right]
= 0
\end{equation}

\noindent where $T_{\rm ex}$ represents the excitation temperature of the transition.
Values of $T_{\rm ex}$ satisfying this condition are turning points in the column
density equation. Using the Turner (\cite{turner91}) high temperature approximations for
the partition function $Q(T_{\rm ex})$, it is easy to show that these turning points are
minima (as the second derivative is positive) and also that they occur at values of 
$T_{\rm ex} = E_{\rm u}/k$ for linear molecules and $T_{\rm ex} = 2\,E_{\rm u}/3\,k$ for
symmetric and asymmetric top molecules. The resulting equations for the minimum
beam-averaged column
density may thus be written as:

\begin{eqnarray}
\label{eqn:nmin_lin}
N_{\rm min} & = & \frac{3k}{8\pi^{3}}\,\frac{\int T_{R} {\rm d}v}{\nu
S\mu^{2}g_{I}g_{K}}\,Q\left(E_{\rm u}/k\right)\, e \\
\label{eqn:nmin_asymm}
N_{\rm min} & = & \frac{3k}{8\pi^{3}}\,\frac{\int T_{R} {\rm d}v}{\nu
S\mu^{2}g_{I}g_{K}}\,Q\left(2E_{\rm u}/3k\right)\, e^{3/2} 
\end{eqnarray}

\noindent where Eq.~\ref{eqn:nmin_lin} is used for linear molecules such as CO and
HC$_{3}$N and Eq.~\ref{eqn:nmin_asymm} is appropriate for symmetric and
asymmetric top molecules such as CH$_{3}$CN and H$_{2}$CS.

We used Equations \ref{eqn:nmin_lin} and \ref{eqn:nmin_asymm} to determine the
beam-averaged column
density lower limits ($N_{\rm min}$) for the lines detected in the survey. Values of 
$N_{\rm min}$ 
are shown in Table \ref{tbl:nmin}. 
Self-absorbed and blended lines were excluded from the analysis. 

\begin{table}[ht]
\caption{Lower limits to column density for the molecular species
 detected in the survey.}                                       
\label{tbl:nmin}
\begin{tabular}{ll}\hline
Species & N$_{\rm min}$ (cm$^{-2}$) \\\hline
$^{13}$CO & 1.9\,\,10$^{16}$ \\
C$^{17}$O & 1.5\,\,$10^{15}$ \\
SO & 2.7\,\,10$^{13}$ \\
HCS$^{+}$ & 1.6\,\,10$^{12}$ \\
CS & 1.7\,\,10$^{13}$ \\
H$^{13}$CN & 1.4\,\,10$^{12}$ \\
H$^{13}$CO$^{+}$ & 5.1\,\,10$^{11}$ \\
H$_{2}$CS & 1.8\,\,10$^{13}$ \\
C$_{2}$H & 1.6\,\,10$^{14}$ \\
H$_{2}$CO & 4.1\,\,10$^{13}$ \\\hline
\end{tabular}
\end{table}

\subsection{Upper limits to column density for undetected species}
\label{sect:uplim}

The column density equation (Eq.~\ref{eqn:coldensity}) may be used to determine an
upper limit to the beam-averaged column density of molecular species that were not detected in the
survey (e.g.~Hatchell et al.~\cite{htmm98a}). In this case assumed values for the maximum integrated intensity of an
undetected line and the excitation temperature of the gas must be used in place of $\int
T_{R}\,\,{\rm d}v$ and $T_{\rm rot}$. The
maximum integrated intensity for an undetected line is estimated from a 3$\sigma$
upper limit to the receiver temperature and an assumed value for the linewidth. 

For all
the upper limits derived in this paper we take the assumed linewidth to be the average
linewidth of the detected lines, i.e. 4 MHz or 3.5 km\,s$^{-1}$ at 345 GHz. We calculate upper limits to
column density for two assumed temperatures (20 and 40 K), to bracket the rotation
temperature of $\sim$ 20 K as derived from the methanol rotation diagrams and also the
values derived by Molinari et al.~(\cite{mol96}, \cite{mol98b}) from ammonia and
millimetre-wave dust
continuum observations (27 and 40 K respectively). It should be noted that the latter
temperature estimate from Molinari et al.~(\cite{mol98b}) is of the temperature at the
outer radius of the millimetre-wave core. 

Upper column density limits were calculated for various astrochemically important
species that have been observed towards hot molecular cores and ultracompact HII
regions, so that we may investigate the evolution of the molecular gas towards these
later stages. We also selected species that are contained in chemical models of these
regions (e.g.~Viti \& Williams \cite{vw99}). The  
upper column density limits are given in Table \ref{tbl:nmax}, along with the line
used for the calculation and its rest frequency.

\begin{table*}[ht!]
\caption{Upper limits to column density for lines that were not detected in the survey.
The line used for the calculation is indicated, along with its rest frequency.}                                       
\label{tbl:nmax}
\begin{tabular}{lllll}\hline
Species & Line & $\nu$(rest) & \multicolumn{2}{c}{N$_{\rm max}$ (cm$^{-2}$)}  \\
 & & (GHz) & $T_{\rm ex} = 20$ K & $T_{\rm ex} = 40$ K\\ \hline
CH$_{3}$CN & 18(0)--17(0) & 331.072 & 2.9\,\,10$^{15}$ & 1.9\,\,10$^{14}$ \\
CH$_{3}$CCH & 20(0)--19(0) & 341.741 & 1.0\,\,10$^{16}$ & 3.7\,\,10$^{14}$\\
HCOOCH$_{3}$ & 17(5,12)--16(4,13) A & 343.150 & 7.9\,\,10$^{14}$ & 2.3\,\,10$^{14}$\\
HC$_{3}$N & 38--37 & 345.609 & 1.9\,\,10$^{18}$ & 1.2\,\,10$^{15}$ \\
SO$_{2}$ & 4(3,1)--3(2,2) & 332.505 & 3.6\,\,10$^{13}$ & 4.6\,\,10$^{13}$\\
OCS & 28--27 & 340.449 & 1.5\,\,10$^{18}$ & 7.8\,\,10$^{15}$\\ \hline
\end{tabular}
\end{table*}

It can be seen from Table \ref{tbl:nmax} that upper limits for species with only
high-excitation lines ($E_{\rm u}/k \ge 150$ K) are not well constrained for the $T_{\rm
ex} = 20$ K case. This is due to the exponential term in Eq.~\ref{eqn:coldensity},
which significantly raises the column density upper limit for lines possessing large
values of $E_{\rm u}/k$ in relation to the excitation temperature of the gas. As the
assumed excitation temperature rises the derived column density upper limit decreases,
until the point where the partition function term $Q(T_{\rm ex})$ begins to dominate
over the exponential ($T_{\rm ex} \simeq Eu/k$). 

The low temperatures used in the calculation may not be appropriate for species that are
thought to predominantly originate from the evaporation of dust grain ice mantles
(CH$_{3}$CN and HCOOCH$_{3}$). These species, if they are present in the gas phase,
are expected to have  evaporated from the ice mantles at gas kinetic temperatures above
$\sim$ 90 K 
(e.g.~Hatchell et al.~\cite{htmm98a}). Purely gas-phase chemical models predict extremely
low abundances of these molecules (Millar et al.~\cite{mmg97}). In order to
account for the possibility that \object{IRAS 23385+6053} contains a small embedded hot molecular
core heated by the central protostar we have evaluated column density upper limits for
CH$_{3}$CN and HCOOCH$_{3}$ assuming an excitation temperature of 150 K. These limits
are: $N_{\rm max} = 8.8\,\,10^{13}$ for CH$_{3}$CN and $N_{\rm max} = 3.1\,\,10^{14}$ for
HCOOCH$_{3}$. We will investigate the likelihood that \object{IRAS 23385+6053} contains an
embedded hot molecular core and set limits on its possible size in Sect.~\ref{sect:hotcore}. 


\section{Discussion}
\label{sect:discuss}

\subsection{The molecular inventory of \object{IRAS 23385+6053}}

The results of our survey show that \object{IRAS 23385+6053} has a simple molecular inventory. 
The gas is comprised of simple molecules with emission detected from CO, CN, SO,
HCS$^{+}$, CS, HCN, HCO$^{+}$, C$_{2}$H, H$_{2}$CO, CH$_{3}$OH and H$_{2}$CS. The latter
molecule, H$_{2}$CS, is an uncertain identification as several other low-excitation 
lines of this asymmetric top molecule lying in the frequency range of the survey  were
not detected. Only low-excitation lines of these species were detected with values of
$E_{\rm u}/k \le 100$ K. 
To this inventory we may add the following species observed
toward \object{IRAS 23385+6053} by other authors:  NH$_{3}$ (Molinari et al.~\cite{mol96}), SiO
(Molinari et al.~\cite{mol98b}) and CH$_{3}$CCH (Brand et al.~\cite{bcpm01}).

We did not detect these last two species (SiO and CH$_{3}$CCH) in our survey due to a
combination of excitation and the sensitivity of our observations. The lines of these
species in the 330--360 GHz frequency range are of moderate to high excitation; $E_{\rm
u}/k = 75$ K in the case of the SiO 8--7 line and $E_{\rm u}/k = 179$ K for the
CH$_{3}$CCH 20(0)--19(0) line. The upper limit that we derived for CH$_{3}$CCH in
Sect.~\ref{sect:uplim} for an excitation temperature of 40 K is larger than the column
density measured by Brand et al.~(\cite{bcpm01}) for the CH$_{3}$CCH J=6--5 K-ladder
lines by roughly an order of magnitude and thus our observations are consistent with
those of Brand et al.~(\cite{bcpm01}). Molinari et al.(\cite{mol98b}) detected the SiO
2--1 line with a source-integrated  column density of 1.4\,\,10$^{14}$ cm$^{-2}$ for an
assumed excitation temperature of 30 K. Using Eq.~\ref{eqn:coldensity} we calculate
a beam-averaged upper column density limit (based upon our non-detection of the SiO
J=8--7 line at 347.331 GHz) of 1.0\,\,10$^{12}$ cm$^{-2}$. This is inconsistent with the
Molinari et al.~(\cite{mol98b}) detection and indicates that either the assumed
excitation temperature is too large or that the emitting area of the SiO J=8--7 line is
small enough for the beam-dilution in our 14\arcsec\ beam to reduce the emission below
our sensitivity limit (\tastar(rms) = 43 mK).

We did not detect emission from any of the complex organic molecules whose high
abundances are a classic signature of hot molecular cores (e.g.~CH$_{3}$CN, HCOOCH$_{3}$
or C$_{2}$H$_{5}$OH). These molecules are predominantly thought to originate from the
evaporation of dust grain ice mantles and their non-detection towards \object{IRAS 23385+6053} 
indicates that the presence of a hot molecular core within \object{IRAS 23385+6053} is unlikely.
A more sensitive search for the hot core tracer methyl cyanide (CH$_{3}$CN) was
conducted recently by Pankonin et al.~(\cite{pcwb01}) towards a number of candidate
massive protostars. Their non-detection of emission from the CH$_{3}$CN J=12--11
K-ladder to a 1$\sigma$ r.m.s.~level of \tastar\ = 13mK allows us to set an upper limit to the
CH$_{3}$CN column density of 5.0\,\,10$^{12}$ cm$^{-2}$ for an assumed excitation
temperature of 40 K (a factor of $\sim 40$ lower than our limit based on the non-detection
of the J=18--17 K-ladder).

Important molecular lines that fall in the frequency bands 
missing from the survey are the HCO$^{+}$ J=3--2 and HCN J=4--3
lines, plus several lines of SO$_{2}$ and CH$_{3}$OH. For the former two species we have
detected their less common $^{13}$C isotopomers and are thus certain that the more common
isotopomers are present in the molecular gas. The remaining two species are asymmetric
rotors and there are many transitions lying in the frequency bands of the survey. Thus
we are reasonably confident that the missing frequency bands do not affect the final
conclusions of the survey.

The molecular inventory of \object{IRAS 23385+6053} is that of a cold dense molecular core with
depleted abundances of complex molecules and exhibiting emission only  from
low-excitation lines of chemically simple species. In many respects the molecular
inventory of \object{IRAS 23385+6053} shares many characteristics with  the line-poor
ultracompact HII regions observed by Hatchell et al.~(\cite{htmm98a}). Both types of
object exhibit emission from low excitation lines with $E_{\rm u}/k \le 100$ K and show
no evidence for hot, dense molecular gas. If IRAS 23385 can be thought of as a
representative massive protostar this may indicate that the molecular inventory of
massive star-forming regions evolves from line-poor protostars  to line-rich hot cores
as the molecular ices are evaporated from dust grains, then back to line-poor
ultracompact HII regions as the chemically rich dense gas of the hot core is dispersed
by the radiation from the newly-born massive stars.

In summary, the molecular inventory of \object{IRAS 23385+6053} suggests that the molecular gas
is cold and dense, with depleted abundances of the more complex saturated species that
trace hot molecular cores. The composition of the gas
appears to be similar to the line-poor ultracompact HII regions observed by Hatchell et
al.~(\cite{htmm98a}), whose limited molecular inventories are ascribed to the lack of a
hot dense molecular core. We investigate this possibility in the following section.

\subsection{Could \object{IRAS 23385+6053} contain a hot core?}
\label{sect:hotcore}

\object{IRAS 23385+6053} does not exhibit any of the classic signs of a hot molecular core,
i.e.~high excitation lines or high abundances of saturated molecules such as CH$_{3}$CN
or HCOOCH$_{3}$. The highest excitation lines detected in our survey have values of
$E_{\rm u}/k \le 100$ K and in the previous section, using the observations of Pankonin et al.~(\cite{pcwb01}),
we set an upper limit for the beam-averaged  CH$_{3}$CN column density of 5.0\,\,10$^{12}$
cm$^{-2}$, which assumes an excitation temperature of 40 K. Nevertheless, given the
small filling factor of the millimetre core (Molinari et al.~\cite{mol98b}) and the beams
of our and Pankonin et al's study it is possible that a small, optically thick hot
molecular core may lie below our sensitivity limits.

To explore this possibility and set an upper limit on the size of any hot molecular core
we modelled the expected CH$_{3}$CN emission from a compact molecular core using the LTE
technique described by Hatchell et al.~(\cite{htmm98a}). This technique assumes that the
emission is in thermal equilibrium, but does not assume optically thin emission. The
density of the \object{IRAS 23385+6053} millimetre core is $\sim$ 10$^{7}$ cm$^{-3}$ (Molinari
et al.~\cite{mol98b}) and this is more than sufficient to thermalise the CH$_{3}$CN
transitions. The LTE analysis technique predicts the line receiver temperature \trstar\
as a function of a column density $N$, the gas kinetic temperature $T_{\rm kin}$ and the
angular source FWHM $\Theta_{\rm s}$. Here, as we can only assume an upper limit for the
observed line receiver temperature, we do not have sufficient information to
extract either the kinetic temperature or column density and so we model the emission
based on assumed values of the kinetic temperature and the CH$_{3}$CN source-averaged 
column density.
 
We assume that as the gas is thermalised  $T_{\rm kin} \simeq T_{\rm ex}$. We use an
upper limit for the line receiver temperature \trstar\ of 3$\sigma$, which is 51 mK in
the Pankonin et al.~(\cite{pcwb01}) study for the J=12--11 K-ladder and 116 mK for the
CH$_{3}$CN J=19--18 ladder that was undetected in our survey. Within both ladders we
selected the K=0 component for our LTE calculations. To calculate the CH$_{3}$CN
source-averaged column density we assume a standard hot core abundance for CH$_{3}$CN of
10$^{-8}$ relative to H$_{2}$ and use density and radius values for the millimetre core
of $\sim 10^{7}$ cm$^{-3}$ and 0.048 pc as derived from the envelope model of Molinari
et al.~(\cite{mol98b}). The LTE models show that as the H$_{2}$ density of the millimetre 
core is high, the CH$_{3}$CN emission is extremely optically thick for both the J=12--11
and J=19--18 K-ladders and remains so even for high gas temperatures, small core radii
and low CH$_{3}$CN abundances.  In the optically thick case, the predicted receiver
temperatures are practically independent of the column density or abundance of
CH$_{3}$CN and depend solely on the kinetic temperature and beam filling factor of the
core (see Equation 1 of Hatchell et al.~\cite{htmm98b}).

For a kinetic temperature of 150 K we determine a maximum hot core angular diameter of
0\farcs7, using Pankonin et al's detection limit and their FWHM beamwidth of
36\arcsec. The angular size of an optically thick core is inversely proportional to the 
square root of the kinetic temperature, so there is a slight increase in the maximum
core size for colder cores (0\farcs8 for a 100 K core) and a corresponding decrease
for hotter cores (0\farcs5 for a 300 K core). Even though our survey is a factor of
two less sensitive than the work of Pankonin et al.~(\cite{pcwb01}), we obtain tighter
constraints upon the maximum hot core angular diameter because our FWHM beamwidth
(14\arcsec) is less than half that of Pankonin et al. For a \trstar\ 3$\sigma$ 
upper limit of 112 mK we determine maximum hot core angular diameters of 0\farcs5,
0\farcs4 and 0\farcs3 for kinetic temperatures of 100, 150 and 300 K respectively.

We have taken a kinetic temperature of 100 K as a lower limit to the temperature of a
hot molecular core. The high abundances of saturated molecules that are present in hot
cores are predominantly thought to arise from the evaporation of dust grain ice mantles
(e.g.~Brown et al.~\cite{bcm88}), and these mantles would remain in the
solid phase at temperatures much below 100 K. Thus, the maximum angular size of any hot
molecular core that may be associated with \object{IRAS 23385+6053} is 0\farcs5, which
corresponds to a spatial radius of 0.006 pc at the assumed distance to \object{IRAS 23385+6053}
of 4.9 kpc. This is roughly an eighth of the radius that Molinari et al.~(\cite{mol98b})
derive for the \object{IRAS 23385+6053} millimetre core. The radii of hot molecular cores known
to be associated with ultracompact HII regions has been measured as typically 0.03--0.06 pc
via similar LTE modelling (Hatchell et al.~\cite{htmm98a}) or interferometric CH$_{3}$CN
observations Olmi et al.~\cite{ocnw96}; Hofner et al.~\cite{hkcwc96}). 

Using a simple argument based upon the Stefan-Boltzmann law we can estimate the
minimum angular size of a hot core that is consistent with the bolometric luminosity of
\object{IRAS 23385+6053}. The equilibrium temperature $T_{\rm eq}$ of the core is
inversely  proportional to the angular radius $\theta$ of the core (strictly, $T_{\rm
eq}^{\,\,\,4} \propto \tan^{-2}\,\theta$) and so increasing the equilibrium temperature of
the core decreases its angular radius. Assuming that the majority of the bolometric
luminosity originates from the hot core and that the bolometric luminosity and distance of
the core are 1.6\,\,10$^{4}$ \lsol\ and 4.9 kpc respectively (Molinari et
al.~\cite{mol98b}), we derive a minimum angular diameter for a 100 K equilibrium temperature
core of 2\farcs8. The validity of this approach may be checked by comparing the predicted
equilibrium temperature of the IRAS 23385+6053 millimetre-wave core to that determined by
Molinari et al.~(\cite{mol98b}) from an envelope radiative transfer model. Using a core
angular radius of 2\arcsec\ the predicted equilibrium temperature is 60 K, fully
consistent with the temperature of 40 K determined by Molinari et al.~(note that their
temperature is that at the external radius of the core and increases towards the centre).

The maximum possible hot core diameter for T $=$ 100 K that we have determined from the
CH$_{3}$CN upper limit (0\farcs5) is considerably smaller than the minimum possible
diameter derived from the simple Stefan-Boltzmann luminosity argument (2\farcs8). Only by
increasing the temperature of the core to greater than 300 K can we reconcile the
CH$_{3}$CN non-detection and the Stefan-Boltzmann predicted core size. However, such a
high-temperature core is inconsistent with the non-detection of IRAS 23385+6053 at 15
$\mu$m (Molinari et al.~\cite{mol98b}) for all but extremely low-mass cores. We thus
conclude that IRAS 23385+6053 is not associated with a hot molecular core.

The only supporting evidence for hot molecular gas towards \object{IRAS 23385+6053} that
would strengthen the case for a hot core is the association with an H$_{2}$O maser
(Molinari et al.~\cite{mol96}). The collisional pumping of H$_{2}$O masers  requires the
presence of gas densities $\sim10^{7}$ cm$^{-3}$ and kinetic temperatures of several
hundred K (Elitzur et al.~\cite{ehm89}). However, the positional accuracy of
the  H$_{2}$O maser emission is only good to within $\sim 1$\arcmin\ (it has only been
observed to date as part of the Medicina H$_{2}$O maser survey, Valdettaro et
al.\cite{h2omedicina}). Thus the maser is as likely to originate from shocks within the
outflow  associated with \object{IRAS 23385+6053} than within a possible  hot molecular
core and by itself  does not provide conclusive evidence for the presence of a hot
molecular core.  

IRAS 23385+6053 is not associated with either a hot molecular core or an ultracompact HII
region. There is thus a strong possibility that it is in an evolutionary phase prior to
these two phenomena. We will explore the nature of IRAS 23385+6053 in
the next section, paying particular attention to the constraints that chemistry may place
upon its evolutionary phase.

\subsection{The physical and chemical nature of \object{IRAS 23385+6053}}

Our molecular line survey reveals that \object{IRAS 23385+6053} is comprised of cold gas with a
molecular inventory limited to simple species such as CO, CS, SO, HCN and CH$_{3}$OH.
We do not detect emission from highly excited molecular lines (all lines detected in the
survey have values of $E_{\rm u}/k \le 100$ K), nor lines from saturated molecules
believed to arise in the gas-phase from evaporation of dust grain molecular ice mantles.
As a result it is unlikely that \object{IRAS 23385+6053} is associated with a hot molecular core,
although our data do not preclude the existence of a small hot core with a radius less
than 0.006 pc.

A natural supposition given the non-detection of evaporated molecular species, the gas
temperature of $\sim 20$K determined from the methanol rotation diagrams and the
candidate protostellar nature of \object{IRAS 23385+6053} is that the molecular gas is in a cold,
depleted ``pre-switch on'' state. In this scenario the massive protostar at the heart of
the \object{IRAS 23385+6053} millimetre-wave core has either not yet begun to warm the surrounding gas
or has not yet warmed a sufficient volume of gas to be detectable. Thus many molecular
species could be currently frozen out in dust grain ice mantles. 

The temperature of the
gas as probed by methanol in this study, CH$_{3}$CCH in the study of Brand et
al.~(\cite{bcpm01}) and the NH$_{3}$ study of Molinari et al.~(\cite{mol96}) support this
hypothesis, with gas temperatures ranging from 20--40 K. The small linewidths seen in our
survey show that the gas is relatively quiescent compared to ultracompact HII regions and
hot molecular cores which typically exhibit much broader linewidths of around 8--10
km\,s$^{-1}$(Hatchell et
al.~\cite{htmm98a}). \object{IRAS 23385+6053} is revealed as a cold, dense, relatively quiescent
molecular core, displaying protostellar hallmarks such as a strong outflow and high
sub-millimetre to bolometric luminosity ratio (Molinari et al.~\cite{mol98b}).

Does the chemistry of \object{IRAS 23385+6053} also support this hypothesis? Recent chemical models
(Rodgers \& Charnley \cite{rc03}; Viti \& Williams \cite{vw99}) investigate the
time-dependent evaporation of molecular ices within massive star-forming cores and their
subsequent chemical evolution. The models predict that certain molecular species will be
evaporated from the grain mantles at different epochs, depending upon their binding
energy to the grain surface and the rise in luminosity of the central protostar or YSO.
For example, in the collapse model of Rodgers \& Charnley (\cite{rc03}) the abundance of
SO$_{2}$ is predicted to peak later than SO.

Before contrasting the column density ratios of various species to those predicted by
the models it is useful to stress that due to the paucity of molecular lines detected in
the survey we are able to provide stringent column density limits for only one species
(methanol). The column density upper and lower limits that we have derived are all
beam-averaged across a 14\arcsec\ FWHM beam and it is not at all certain that the
molecular species  trace the same emitting area of gas. The high critical density lines
of CH$_{3}$CN with its large dipole moment are unlikely to trace the same gas as, for
example, the low critical density lines of CH$_{3}$OH with its relatively low dipole
moment. With these caveats in mind, it is nevertheless instructive to compare the
observed chemistry of \object{IRAS 23385+6053} to that predicted by molecular evaporation
chemical models in order to determine whether the models can constrain the evolutionary
state of \object{IRAS 23385+6053}.

We have chosen to calculate the column density ratios as a fraction of the methanol
column density so that the resulting ratio is either an upper or lower limit depending
upon whether the other species is a detection or a non-detection. Column density ratios
were calculated for those species modelled by Rodgers \& Charnley (\cite{rc03}) and Viti
\& Williams (\cite{vw99}) and are found in Table \ref{tbl:Nratio}. 
In order to calculate
the column densities of more common isotopomers from their less common variants
(e.g.~HCO$^{+}$ from H$^{13}$CO$^{+}$) we assumed that the more common isotopomer was 
optically thin and used standard interstellar isotope ratios. 

For the undetected species with upper limits to column density we used values
assuming an excitation temperature of 20 K to maintain consistency with the rotation
temperature drived from the methanol rotation diagrams. This may be inappropriate for
the hot core tracer molecules HCOOCH$_{3}$ and CH$_{3}$CN, but as their column density
upper limits  slowly decrease with temperature until $T_{\rm ex} \simeq E_{\rm u/k}$ the
minimum column density ratios quoted in Table \ref{tbl:Nratio} are still valid for
excitation temperatures less than this value.

\begin{table}[ht!]
\caption{Column density ratios between species modelled in the evaporation models of 
 Rodgers \& Charnley (\cite{rc03}) and Viti \& Williams (\cite{vw99}).Upper and lower
 limits to the ratios are indicated.}                                       
\label{tbl:Nratio}
\begin{tabular}{ll}\hline
Species & $N/N_{CH_{3}OH}$  \\ \hline
SO & $> 0.03$ \\ 
CS & $> 1.5$ \\
HCO$^{+}$ & $> 0.03$ \\ 
H$_{2}$CS & $> 0.02$ \\
H$_{2}$CO & $>0.04$ \\
CH$_{3}$CN & $ < 0.01$ \\
HCOOCH$_{3}$ & $< 0.8$ \\
SO$_{2}$  & $< 0.04$ \\
OCS & $<1500$ \\ \hline
\end{tabular}
\end{table}

The chemical models of Rodgers \& Charnley (\cite{rc03}) and Viti \& Williams
(\cite{vw99}) were not explicitly calculated for the \object{IRAS 23385+6053} molecular core and
so we cannot constrain the chemical timescale in an absolute sense.  However, Viti \&
Williams (\cite{vw99}) showed that whilst the evaporation timescale depends upon  the
rapidity of the molecular ice evaporation, the subsequent gas-phase evolution is similar
for all evaporation models. We can thus use  the column density ratios in Table
\ref{tbl:Nratio} to determine if the chemistry of IRAS 23385+6053 is consistent with the
early, middle or late phases predicted by the chemical models. These three phases
represent the early stages of the evaporation in which the chemistry is dominated by
small simple molecules with low surface binding energies (e.g.~CO and CH$_{4}$), the
middle phase when most species have just evaporated from the grains and the late phase
where the chemistry is dominated by the gas-phase reactions between evaporated species. 

Broadly speaking the chemistry of \object{IRAS 23385+6053} is consistent with the 
middle evaporation phases predicted by both models. In the model of Viti \& Williams (\cite{vw99})
methanol comes off the grains roughly in the middle of the evaporation process, preceded
by simple molecules with low surface binding energies and followed by the more complex
species such as HCOOCH$_{3}$, CH$_{3}$CO and C$_{2}$H$_{5}$OH. The abundance of methanol
after the initial evaporation  rises rapidly, with a plateau at an abundance of $\sim
10^{-11}$  which lasts for a few 10$^{3}$ years followed by a second rapid rise to a
final plateau at an abundance of $\sim 10^{-8}$. 

The observed column density ratios in Table \ref{tbl:Nratio} show that methanol is more
abundant than almost every other species with the exception of OCS and CS. The ratios for
SO, CS, HCO$^{+}$,  H$_{2}$CS and H$_{2}$CO are all lower limits and it is possible that
these species are more abundant than methanol. Based upon the assumed excitation
temperatures used in the calculation of the column density lower limit and the estimated
kinetic temperature of the gas from the methanol rotation diagrams we expect the true
column density to be underestimated by at most a factor of 10 (e.g.~Thompson et
al.~\cite{tmm99}). For the ratios that are upper limits (CH$_{3}$CN, HCOOCH$_{3}$, SO$_{2}$
and OCS) we can be reasonably certain that the only species that may be more abundant than
methanol is OCS. 

Methanol is predicted in the model of Viti \& Williams (\cite{vw99})  to be more
abundant than almost every other molecule in Table \ref{tbl:Nratio} during the early
evaporation phase. The exceptions to this rule are OCS which is more abundant than
methanol at all times, CS which has roughly the same abundance during its early
evaporation, H$_{2}$CO which is always more abundant than methanol and CH$_{3}$CN which
evaporates from grains marginally  before methanol and is more abundant until methanol
evaporates. In the later phases many species are predicted to have larger abundances than
methanol, in particular CS, H$_{2}$CS, SO, and the aforementioned OCS and H$_{2}$CO.
The models of Rodgers \& Charnley (\cite{rc03}) broadly agree with these predictions for
the late phase; they do not calculate abundance ratios for the early ice evaporation
phase  (note that their Figure 11 begins at a time 100 years \emph{after} the evaporation
phase).

On balance the chemistry of \object{IRAS 23385+6053} resembles that predicted by Viti \& Williams
(\cite{vw99}) for the middle ice evaporation phase immediately prior to the development
of a hot molecular core. The abundance of SO rises faster than that of methanol after
the middle phase. HCO$^{+}$ is more abundant than methanol during the early phase of
evaporation and less abundant in the middle and later phases. The SO$_{2}$ abundance is
lower than that of methanol in the early and middle stages, rising toward that of
methanol in the late stages. These results constrain the chemical timescale of IRAS
23385+6053 to that of the middle phase, i.e.~where the majority of species are beginning
to be evaporated from the dust grain ice mantles.

The major inconsistencies that arise between the observed ratios and the model
predictions are for CS and OCS, where the observed ratios are larger than the model
predictions. It is possible that the CS abundance is severely  underestimated due to
high optical depth. We can discount this possibility due to the non-detection of the
C$^{34}$S line at 337.397 GHz, which allows us to set a maximum optical depth of 2.6 for
the main isotopomer. The column density upper limit for OCS cannot be more accurately
constrained as the only OCS lines lying in our survey range are of high excitation, 
possessing values of $E_{\rm u}/k \simeq 250$ K.

Both the physical and chemical properties of \object{IRAS 23385+6053} point towards its
protostellar nature and identify the \object{IRAS 23385+6053} millimetre core as being on the
verge of developing into a hot molecular core. However,  in order to confirm this
hypothesis the chemical abundances of the molecular gas must be constrained by further
observations of lower excitation lines and more detailed modelling. In particular as the
molecular core is unresolved by our survey observations the relative emitting sizes and
beam filling factors of the various species are extremely uncertain. Interferometric
observations are a priority to try and resolve the emission and derive firm abundance
limits for the chemistry of \object{IRAS 23385+6053}. 


\section{Summary and conclusions}

We have carried out a molecular line survey with a total frequency range of 27.2 GHz  of
the candidate massive Class 0 protostar \object{IRAS 23385+6053}. We detected emission from 27
lines originating from 11 molecular species. One  line feature could not be identified
with any known lines in the JPL Molecular Spectroscopy Database or with other
observational line lists (Jewell et al.~(\cite{jhls89}); Macdonald et
al.~(\cite{mghm96}); Schilke et al.~(\cite{sgbp97});  Thompson \& Macdonald
(\cite{tm99})). No emission was detected from high excitation lines ($E_{\rm u}/k \ge
100$ K) or from the complex saturated molecules (e.g.~CH$_{3}$CN, HCOOCH$_{3}$ or
C$_{2}$H$_{5}$OH)  observed toward hot molecular cores. Over a third of the identified lines
originate from methanol (CH$_{3}$OH).

We derived the rotation temperature and column density for methanol using the rotation
diagram approach and estimate lower beam-averaged column density limits for molecular
species with one or two line detections (e.g.~Thompson \& Macdonald \cite{tm99}). Upper
limits for undetected species were determined from the r.m.s.~noise level of appropriate
spectra. We draw the following conclusions from our survey:

\begin{enumerate}

\item The molecular inventory of \object{IRAS 23385+6053} resembles that of the line-poor
ultracompact HII regions observed by Hatchell et al.~(\cite{htmm98a}). They both exhibit
emission from low excitation transitions of simple molecules and show no signs of hot
core emission. Given the supposed massive protostellar nature of \object{IRAS 23385+6053} this
may indicate that the molecular inventory of massive star-forming regions evolves from
line-poor protostellar cores through a line-rich hot molecular core phase and back to a
line-poor ultracompact HII region phase when the hot dense molecular core has been
dispersed.

\item We see no evidence of a hot molecular core associated with \object{IRAS
23385+6053}.  No emission was detected from high-excitation lines that are tracers of hot
molecular gas and the rotation temperature derived from the methanol emission is $\sim 20$
K.  We rule out the presence of a hot core by a combination of the LTE modelling of the
CH$_{3}$CN emission that is consistent with our non-detection of the CH$_{3}$CN J=19--18
K-ladder and a simple argument based upon the Stefan-Boltzmann law and the bolometric
luminosity of any possible hot molecular core.  The H$_{2}$O maser associated with
\object{IRAS 23385+6053}  (Molinari et al.~\cite{mol96}) must be associated with outflow
shocks rather than the dense molecular core.

\item The chemical composition of the \object{IRAS 23385+6053} molecular core is consistent with
the predictions of Viti \& Williams (\cite{vw99}) for a molecular core in the middle
evaporation phase, i.e.~when the majority of species are beginning to be 
evaporated from dust grain ice mantles. This confirms the hypothesis of Molinari et
al.~(\cite{mol98b}) that \object{IRAS 23385+6053} is an extremely young massive protostellar
object, possibly on the verge of developing into a hot molecular core. 

\end{enumerate}

\begin{acknowledgements} The authors would like to thank Neil Alvey and Samantha Large for
their assistance with the data reduction and line identifications. We would also like to
thank the referee, Todd Hunter, for providing several useful suggestions which
considerably improved this paper, particularly the argument for the lack of a hot
molecular core. MAT is supported by a PPARC postdoctoral grant. This research has made use
of the  SIMBAD astronomical database service operated at CDS, Strasbourg, France  and the
NASA Astrophysics Data System Bibliographic Services. Quicklook 2MASS images were obtained
as part of the Two Micron All Sky Survey (2MASS), a joint project of the University of
Massachusetts and the Infrared Processing and Analysis Center/California Institute of
Technology, funded by the National Aeronautics and Space Administration and the National
Science Foundation. MSX 8 $\mu$m data were obtained from the NASA/IPAC Infrared Science
Archive, which is operated by the Jet Propulsion Laboratory, California Institute of
Technology, under contract with the National Aeronautics and Space Administration.
\end{acknowledgements}

\end{document}